\documentclass[10pt,conference]{IEEEtran}
\usepackage{hyperref}
\usepackage{amsmath,amssymb,amsfonts}
\usepackage{color}
\newcommand{\zhang}[1]{\textcolor{red}{\textbf{Min:\ }#1}}

\usepackage{url}
\usepackage{txfonts}
\usepackage{balance}
\usepackage{multirow}
\usepackage{verbatim}

\usepackage{multirow}
\newcommand{\klabel}[1]{[\textcolor{blue}{\textsf{#1}}]}
\usepackage{listings}
\newcommand{\K}{\mathbb{K}}
\newcommand{\rust}{ {\tt Rust}}
\newcommand{\krust}{ {\tt KRust}}
\newcommand{\ds}{``$\cdots$'' }
\newcommand{\LL}{\left\langle }
\newcommand{\RR}{\right\rangle }
\newcommand*{\bfrac}[2]{\genfrac{\lbrace}{\rbrace}{0pt}{}{#1}{#2}}
\newcommand{\z}{\left\langle}
\newcommand{\y}{\right\rangle}

\begin{document}

\title{$\krust$: A Formal Executable Semantics of Rust}

\author{
    \IEEEauthorblockN{
        Feng Wang\IEEEauthorrefmark{1},
        Fu Song\IEEEauthorrefmark{1},
        Min Zhang\IEEEauthorrefmark{2},
        Xiaoran Zhu\IEEEauthorrefmark{2} and
        Jun Zhang\IEEEauthorrefmark{1}
    }
    \IEEEauthorblockA{\IEEEauthorrefmark{1}School of Information Science and Technology, ShanghaiTech University, Shanghai, China}
    \IEEEauthorblockA{\IEEEauthorrefmark{2}Shanghai Key Laboratory of Trustworthy Computing, East China Normal University, Shanghai, China}
}

\maketitle

\begin{abstract}
Rust is a new and promising high-level system programming language. It
provides both memory safety and thread safety through its novel mechanisms such as ownership, moves and borrows. Ownership system ensures that at any point there is only one owner of any given resource. The ownership of a resource can be moved or borrowed according to the lifetimes. The ownership system establishes a clear lifetime for each value and hence does not necessarily need garbage collection.
These novel features bring Rust high performance, fine low-level control of C and C++, and unnecessity in garbage collection,  which differ Rust from other existing prevalent languages. For formal analysis of Rust programs and helping programmers learn its new mechanisms and features, a formal semantics of Rust is desired and useful as a fundament for developing related tools.
In this paper, we present a formal executable operational semantics of a realistic subset of Rust, called $\krust$.
The semantics is defined in $\K$, a rewriting-based executable semantic framework for programming languages.
The executable semantics yields automatically a formal interpreter and verification tools for Rust programs.
$\krust$ has been thoroughly validated by testing with hundreds of tests, including the official Rust test suite.
\end{abstract}


\section{Introduction}


%

Recently, a new system programming language Rust was designed and implemented by Mozilla \cite{matsakis2014rust}, aiming at achieving both high-level safety and low-level control in developing system software, such as operating systems, device drivers, game engines and web browsers. Like most modern high-level languages, Rust guarantees memory safety and thread safety, and meanwhile it supports zero-cost abstractions for many common system programming idioms and provides fine low-level control over the use of memory, without needing a garbage collector.
One key to meeting all these promises is Rust's novel system of ownership, moves and borrows.
The ownership system establishes a clear lifetime for each value, making garbage collection
unnecessary in the core language. Moreover, it also prevents data-races at compile-time. The three mechanisms are checked at compile time and carefully designed to complement its static linear type system.
Rust has been used to implement operating system \cite{Redox}, parallel browser engine \cite{ABGMMMS16}, Intel SGX Enclave \cite{DDLCZCWW17}, etc. (organizations running Rust in production refers to \footnote{\url{https://www.rust-lang.org/en-US/friends.html}.})

The new features make the semantics of Rust different from other common languages, which brings new difficulties in reasoning about Rust programs.
 The difficulty constantly baffles developers and has become a common topic of question-and-answer websites (for instance, \cite{Stackoverflow}).
The ownership of an object is a variable binding. When a variable goes out of its scope, Rust will free the bound resources.
The ownership of an object can be \emph{transferred}, after which the object cannot be accessed via the outdated owner.
This ensures that there is exactly one binding to any object.
Instead of transferring ownership, Rust provides borrows which allows an object to be shared by multiple references.
There are two kinds of borrows in Rust: \emph{mutable references} and \emph{immutable references}.
A mutable reference can mutate the borrowed object if the object is mutable, while an immutable reference cannot mutate the borrowed object even the object itself is mutable.
The basic ownership discipline enforced by Rust is that an object shared by multiple references is immutable. This property
eliminates a wide range of common low-level programming errors, such as use-after-free, data races, and iterator invalidation.
These specific semantic rules are unusual, and hence the semantics of other modern programming languages such as C/C++, Java and Javascript cannot be directly adapted to Rust.

To remedy this situation, a formal semantics of Rust is desired and useful as a fundament for reasoning about Rust
programs in a formal way and developing related computer-aided tools. To our knowledge, the formal semantics of Rust has not
yet been well studied, which impedes further developments of formal analysis and verification tools for Rust programs.
In this paper, we make a major step toward rectifying this situation by giving the first formal operational semantics of a realistic subset of Rust.
We design a formal operational semantics of Rust capturing ownership, ownership moves and borrows. To avoid our semantics to be just ``paper work",
we formalize the semantics in $\K$ framework \cite{ro2010} (\url{http://kframework.org}), a scalable semantic framework for programming languages which has been successfully applied to C \cite{ER12} and Java \cite{BR15}.
We call the $\K$ definition of the semantics \krust. \footnote{$\krust$ and all the related sample examples are available for downloading from:		\url{http://sist.shanghaitech.edu.cn/faculty/songfu/Projects/KRust}.
} {\bf To the best of our knowledge, it is the first formal executable semantics for Rust}.

 There are several benefits from the formal semantics defined in $\K$ framework.
 Firstly, the semantics defined in $\K$ is both machine readable and executable, from which an interpreter of Rust is generated automatically. Being executable, the semantics has been thoroughly tested with hundreds of tests, including the official Rust test suite. Secondly, $\K$ provides a simple notation for modular semantics of languages, making the semantics easy to define and extensible.
 The semantics offers a formal reference and a correctness definition for implementers of tools such as parsers, compilers, interpreters and debuggers, which would greatly facilitate developers' understanding, freeing them from lengthy, ambiguous, elusive Rust documentations. Moreover, the semantics could  automatically yield formal analysis tools such as state-space explorer for reachability, model-checker, symbolic
 execution engine, deductive program verifier by the language-independent tools provided in $\K$ \cite{SPYKR16}.


\smallskip
\noindent
{\bf Organization}. Section \ref{sec:tour} gives a brief overview of Rust.
In Section \ref{sec:sem}, after introducing the basic notations of $\K$, we present the formal semantics
$\krust$ of Rust in $\K$. Conformance testing and applications of $\krust$ are described in Section \ref{sec:test}. Section \ref{sec:rel}
discusses related work. Finally, we conclude the work with a discussion in Section \ref{sec:discu}.


\section{A Tour of Rust} \label{sec:tour}
In this section, we give a brief overview of Rust.
Rust is a C-like programming language which contains common constructs from modern programming languages
such as \texttt{if/else} branches, \texttt{while/for} loops, \texttt{functions}, \texttt{compound data structures}, etc.
We will mainly point out distinct features of Rust, compared with other well-known modern programming languages such as C/C++ and Java.

\subsection{Mutability}
Variables are declared using \texttt{let} statements. In default, variable is \emph{immutable}, which means that its value cannot be mutated. 
To declare a mutable variable,  \texttt{mut} is required in the declaration statement.

\begin{lstlisting}
fn main(){
  let x=9;
  x=10; // Error!
  let mut y = 0;
  let mut z: bool;
}
\end{lstlisting}

For instance, the above code declares an immutable variable \texttt{x} at Line 2 and a mutable variable \texttt{y} at Line 4 whose types are inferred at compile-time.
The type can also be explicitly specified in the program like the mutable variable \texttt{z} at Line 5.
The Rust compiler will issue an error at Line 3, as the immutable variable \texttt{x} is reassigned. This is different from other modern programming languages like C/C++ and Java.
The semantic rules of Rust should take into the mutability of variables into account.


\subsection{Functions}
Functions are declared with the keyword \texttt{fn} and each of them should return \emph{exactly one value}.
There are two ways to return a value if the function definition declares a return type. The first one returns a value in the body of the function explicitly using
the \texttt{return} statement, which is similar to existing prevalent languages.
The another one returns the value of the last expression in the body of the function, if there is no explicit return statement.
However, if the function definition does not declare a return type, Rust will implicitly returns the unit type ().
Indeed, a function definition without declaring a return type is just syntactic sugar for the same function definition with return type ().
Furthermore, Rust has no restriction on the order of function definitions, namely that Rust programs can invoke functions which are defined later.
These features are unusual and introduce tricky corner cases.

\begin{lstlisting}
fn foo(x:i32, y:i32) -> i32 {
	 x+y // return x+y;
}
\end{lstlisting}

The above program defines a function \texttt{foo} which takes two 32-bit integers \texttt{x} and \texttt{y} as arguments and returns
\texttt{x+y}. This function behaves the same as the function which replaces the last expression \texttt{x+y} in  \texttt{foo}  with
the return statement \texttt{return x+y;}.

%
%
%
%

\subsection{Ownership}
Ownership is the key feature of Rust, which guarantees memory safety and thread safety without garbage collection.
The basic form of ownership is \emph{exclusive ownership}, namely that each object has a \emph{unique} owner at any time.
This ensures that at most one reference is allowed to mutate a given location.
When an object is created and assigned to a variable \texttt{x}, the variable \texttt{x} becomes the owner of the object.
If the object is reassigned (as well as parameter passing, etc.) to another variable \texttt{y}, the ownership of the object is
\emph{transferred} from the variable \texttt{x} to the variable \texttt{y}, namely that \texttt{y} becomes the owner of the object
and \texttt{x} is not the owner of the object. This is so-called \emph{move semantics} in Rust. This ownership discipline rules out
aliasing entirely, and thus prevents from data race.
Moveover, if the owner of the object goes out of scope, the object is destroyed automatically without garbage collection.
This is implemented by the Rust compiler which inserts a call to a destructor at the end of the owner's scope, called \emph{drop} in Rust.
This ownership discipline enforces automatic memory management and prevents other errors commonplace in low-level pointer-manipulating programs,
like use-after-free or double-free.  To see this
principle in practice, consider the following sample program.

\begin{lstlisting}
struct Point{
   x: i32,
   y: i32,
}
fn main(){
  let p = Point {x:1, y:2};
  {
    let mut q = p; //q becomes the owner
    q.x = 2;
    println!("{}",p.x);  // Error!
  }
  println!("{}",q.x);  // Error!
}
\end{lstlisting}
\texttt{Point} is a compound data structure consisting of two 32-bit integer fields \texttt{x} and \texttt{y}.
At Line 6, a \texttt{Point} object is created and assigned to the mutable variable \texttt{p}.
The Rust compiler will issue an \emph{error} at Line 10 which accesses the \texttt{x} field of the \texttt{Point} object
via the reference \texttt{p}, as the ownership of the \texttt{Point} object was already transferred from \texttt{p} to \texttt{q} at Line 8.
Moreover, the Rust compiler will also issue another \emph{error} at Line 12 which accesses the \texttt{x} field of the \texttt{Point} object
via the reference \texttt{q}, as the owner \texttt{q} went out of its scope and the \texttt{Point} object was already destroyed.
These scenarios rarely happen in existing prevalent programming languages, thus makes the Rust semantics unusual.

%
%
%
%
%
%

\subsection{Borrowing}
The ownership discipline is a fairly straightforward mechanism for guaranteeing memory safety and thread safety.
However, it is also too restrictive to develop industry programs.
%
%
To address this issue, Rust provides a mechanism used to handle references, called \emph{borrowing}.
There are two different borrowing (i.e., reference types) in Rust: \emph{mutable references} and \emph{immutable references}.
A mutable reference grants temporary \emph{exclusive} \emph{read and write} access to the object,
i.e., each object has at most one mutable reference (without any immutable references). This ensures that mutable references are always unique pointers.
A mutable reference can be \emph{reborrowed} to someone else.
Contrary to mutable references, an immutable reference grants temporary \emph{read-only} access to the object and it is allowed that multiple immutable references refer to the same object.
The design of borrowing in Rust also guarantees memory safety and thread safety.

\begin{lstlisting}
fn main() {
   let x1 = 1;
   let p1 = &x1; // x1 is borrowed immutably
   let q1 = &x1; // x1 is borrowed immutably
   *p1 = 2; // Error!
   let y = &mut x1; // Error!

   let mut x2 = 1;
   let p2 = &x2; // x2 is borrowed immutably
   let q2 = &x2; // x2 is borrowed immutably
   *p2 = 2; // Error!

   let mut x3 = 1;
   let p3 = &mut x3; // x3 is borrowed mutably
   x3 = 2; // Error!
   *p3 = 2; // OK!
   let p4 = &mut x3;  // Error!
}
\end{lstlisting}

We can see borrowing in action in the above example.
In this example, the variable \texttt{x1} is \emph{immutable} which is immutably borrowed twice at Lines 3 and 4.
The compiler will issue an \emph{error} at Line 5 which tries to mutate the value of \texttt{x1}.
It also issues another \emph{error} at Line 6, as immutable variable \texttt{x1} cannot be mutably borrowed.
The code at Lines 8-11 shows that the mutable variable \texttt{x2} can also be immutably borrowed multiple times, but cannot be
mutated by an immutable reference. Line 15 demonstrates that the mutable variable \texttt{x3} cannot be mutated via \texttt{x3}
once it is borrowed. Instead, it can be mutated via the mutable reference \texttt{p3} at Line 16. Line 17 shows that
the mutable variable \texttt{x3} cannot be mutably borrowed more than once.

%
%
%
%
%

\subsection{Lifetime}
Borrowing grants \emph{temporary} access to the object.
Rust associates to each reference with a \emph{lifetime} to specify how long is temporary.
Intuitively, lifetimes are effectively just names for scopes somewhere in the program, but they are not same.
The lifetime of a reference should be included in the lifetime of the borrowed variable.
Rust provides a convention so that lifetimes can be elided in general, which is why they did not show up in
the above examples. Rust also supports named lifetimes which helps the Rust compiler to aggressively infer lifetimes and makes sure all borrows are valid.



\begin{lstlisting}
fn main(){
   let mut x ;
   {
     let y = 1;
     x = &y;   // Error!
   }           
   let z = 1;
   let p = &z; // OK!
}
\end{lstlisting}

The above example illustrates intuition behind lifetime.
There is an \emph{error} at Line 5 as the lifetime of \texttt{x} is not included in the lifetime of \texttt{y}.
Instead, it is fine to borrow \texttt{z} at Line 8, as the lifetime of \texttt{p} is included in the lifetime of \texttt{z}.
Therefore,  Rust's variable context is \emph{substructural}.


\begin{table*}[t]
\caption{The syntax of $\krust$}	
	\begin{center}
		\begin{tabular}{ l  | l}
			\hline
			\textbf{Syntax}         & \textbf{Description} \\
			
			\hline \hline
			Id ::= [a-zA-Z\_][a-zA-Z0-9\_]* & Identifier
			\\ \hline
			Type ::= i8 $|$ u8 $|$ i16 $|$ u16 $|$ i32 $|$ u32 $|$ i64 $|$ u64 $|$ f32 $|$ f64 $|$ isize $|$ usize $|$ char \\ 
			\qquad \quad $|$ \&str $|$  bool \ $|$ Id $|$ \ () \ $|$ [Type;Exp] $|$ fn (Types) -$>$ Type \\
			Types ::= Type* & Variable Types
			\\ \hline
			TypedId ::= Id : Type \\
			TypedIds ::= TypedId*
			& Auxiliary types
			\\ \hline
			ConstAndStatic ::= \texttt{const} Id : Type = Exp ;   $|$ \ \texttt{static} \texttt{mut} ? Id : Type = Exp ;\\
			DeclExp ::= \texttt{let} \texttt{mut} ? Id [: Type] ? [= Exp] ? $|$  ConstAndStatic;
			& Variable declaration	
			\\ \hline
			Op ::= ``+'' $|$ ``-'' $|$ ``*'' $|$ ``/'' $|$ ``\%'' $|$ ``$|$'' $|$ ``\&'' $|$ ``$>>$'' $|$ ``$<<$'' $|$ ``$<$'' $|$ ``$<=$'' $|$ ``$>$''  $|$  ``$>=$''  $|$ ``==''  $|$ ``!=''  $|$ ``$||$'' $|$ ``\&\&'' \\
			Exp ::= Int $|$ Bool $|$ Float $|$ String $|$ Char  $|$ Id $|$ *Id $|$ [Exps] $|$ [Exp;Exp] $|$ vec![Exps]  $|$ (Exp) $|$ Exp[Exp]\\
			\quad  \qquad $|$ \{Exp\} $|$ Ref Exp $|$ Id \{ StructValues \}   $|$ Exp(Exp)  $|$ -Exp $|$ ! \ Exp $|$ Exp Op Exp \\		
			Exps ::= Exp*
			& Expressions
			\\ \hline
			AssignOp ::= ``='' $|$ ``+='' $|$ ``-='' $|$ ``*='' $|$ ``/=''  \\
			AssignmentStmt ::= Id AssignOp Exp ; $|$ \ Id[Exp] AssignOp Exp ;   $|$ \ *Id AssignOp Exp; $|$ Id . Id AssignOp Exp ;
			& Assignment statement
			\\ \hline
			If ::= \texttt{if} Exp \texttt{else} ? Block \\
			While ::= \texttt{while} Exp Block \\
			Loop ::= \texttt{loop} Block $|$ If $|$ While
			
			& If, while and loop statement
			\\ \hline
			Block ::= \{ \} $|$ \{ Stmts \} $|$ \{ Stmts Exp \} & Block
			\\ \hline
			Ref ::= \& $|$ \&mut & Two types of references
			\\ \hline
			Struct ::= \texttt{struct} Id \{ TypedIds \} \\
			StructValue ::= Id : Exp \\
			StructValues ::= StructValue* \\
			StructInstance ::= Id \{ StructValues \}
			& Struct
			\\ \hline
			For ::= \texttt{for} Id \texttt{in} Int..Int Block & For statement
			\\ \hline
			Function ::= \texttt{fn} Id (TypedIds) [-$>$ Type] ? Block
			& Function
			\\ \hline
			Stmts ::= DeclExp  $|$ AssignStmt  $|$ Block $|$ Exp ; $|$ return ; $|$ return Exp ;   $|$ Loop ; $|$ Loop $|$ Function $|$ Struct $|$ For
			& Statements
			\\ \hline
		\end{tabular}
	\end{center}
\label{tab:syntax}
\end{table*}

\section{$\krust$: The Formal Semantics of Rust in $\K$}\label{sec:sem}
In this section, we first introduce the basic notations of $\K$, and then define the addressed formal syntax of Rust.
Finally, we define the configurations and formal semantics $\krust$ of Rust in $\K$.

\subsection{$\K$ Framework}
$\K$ Framwork is a rewrite-based executable semantic framework.
Operational semantics of a programming language can be formally defined with the state of an executing program being represented as a configuration and the semantics of each program statement being defined as $\K$ rules. With the defined operational semantics, $\K$ automatically generates an interpreter which can \emph{execute} programs of the language. Besides, $\K$ also provides formal analysis functionalities such as model checking, symbolic execution, and theorem proving \cite{ro2010}. A number of programming languages have been formalized using $\K$, such as C \cite{ER12}, Python \cite{gut13}, PHP \cite{FM14} and Java \cite{BR15}.


In this section, we briefly introduce the mechanism of how to define operational semantics in $\K$ with an example. Basically, the syntax of a language is defined using BNF with semantic attributes, and the operational semantics is defined by a set of $\K$ rules which describe the effect of atomic program statements over $\K$ configurations.
A $\K$ configuration is essentially a nested cell defined in XML style, which specifies a state of an executing program.

For better understanding $\K$, let us consider the following semantic rule for lookup function in Rust, which is used to
lookup the value of a name (e.g., variable) when a statement which contains that name is being executed.
\[
\left\langle \frac{X}{V} \cdots \right\rangle _{{\sf k}}
\left\langle \cdots X \mapsto L \cdots \right\rangle _{{\sf env}} \left\langle \cdots L\mapsto V \cdots \right\rangle _{{\sf store}}
\]

There are three cells related to the lookup function. The cell {\sf k} (i.e., the cell labeled with {\sf k}) is used to store the computations of a program to be executed. In this example,
$X$ is the next expression/statement to evaluate/execute, where $X$ is a $\K$ label representing a program name. Both {\sf env} and {\sf store} cells are Map type to store key-value pairs, where {\sf env} is used to store the mapping from program names to locations in the form of $X\mapsto L$, and  {\sf store} is used to store the mapping from locations to values in the form of $L\mapsto V$.
The dots ``$\cdots$'' in the cells are structural frames, denoting irrelevant pairs or computations.


For instance, given the following two statements:
\begin{lstlisting}
let a = 1;
let b = a;// b = 1	
\end{lstlisting}

Variable \texttt{a} should be first evaluated to 1 using the lookup rule when the second statement is being executed.

\subsection{Syntax}

Table \ref{tab:syntax}  presents the syntax of a realistic subset of Rust defined in $\krust$. The syntax is described by a dialect of Extended Backus-Naur Form (EBNF)
according to the grammar of Rust \cite{Grammar}.  We use two repetition forms in the definition of the syntax, where ``?'' means zero or one repetition. ``*'' means zero or more repetitions. The option is represented through squared brackets [ ... ] followed with ``?''. For instance, in syntax for function declaration,  ``[-$>$ Type] ?'' means that ``-$>$ Type'' maybe present just once, or not at all. Since Rust shares many common conventions with prevalent functional and imperative programming languages, most of its syntax are easy to understand and hence we omit corresponding explanations.
We remark that Table \ref{tab:syntax} is not a full syntax of Rust, e.g., traits and pattern matching are excluded (c.f. Section \ref{sec:discu}).

\subsection{Configuration}
A $\K$ configuration is represented by nested multisets of labeled cells. Figure \ref{fig:conf} shows the 13 main cells in a configuration for the representation of the state of $\krust$ programs.

\begin{figure}[t]
\begin{center}
\[
\LL \begin{array}{c}
\LL K \RR_{\sf k} \LL Map \RR_{\sf env} \LL \LL List \RR_{\sf fstack} \RR_{\sf control}
\LL Map \RR_{\sf genv} \\
\LL Map \RR_{\sf typeEnv} \LL Map \RR_{\sf store}  \LL Map \RR_{\sf mutType} \LL 0  \RR_{\sf nextLoc}\\
\LL Map \RR_{\sf borrow}\LL Map \RR_{\sf ref} \LL Map \RR_{\sf refType}  \LL Map \RR_{\sf moved}
\end{array}
\RR_{\sf T}
\]	
\caption{The $\K$ configuration for the states of $\krust$ programs}
\label{fig:conf}
\end{center}	
\end{figure}

\begin{table*}[t]
	\caption{The partial semantic rules of $\krust$}
	\begin{center}
		\footnotesize{
			\begin{tabular}{l}
				\hline
				\textbf{Variable declaration and assignment} \\
				$\begin{array}{c}
				\mbox{\normalsize $\left\langle \frac{\texttt{let} \ X \ :\ T;}{.} \right\rangle _{\sf k}$}
				\left\{ \begin{array}{lll}
				\left\langle  X \Rightarrow L  \right\rangle _{\sf env} & \left\langle  L \Rightarrow \ \perp  \right\rangle _{\sf store} & \left\langle  L \Rightarrow T  \right\rangle _{\sf typeEnv} \\
				\left\langle  L \Rightarrow 0  \right\rangle _{\sf mutType} &  \left\langle  L \Rightarrow \ \perp  \right\rangle _{\sf borrow} & \left\langle  L \Rightarrow \ \perp  \right\rangle _{\sf ref} \\
				\left\langle  L \Rightarrow \ \perp  \right\rangle _{\sf refType}&
				\left\langle  L \Rightarrow \ 0  \right\rangle _{\sf moved} &
				\z L + 1\y_{\sf nextLoc}
				\end{array} \right\} \\
				\hfill \klabel{Declaration-of-Immutable-Variable}
				\end{array}$
				
				\hfill
				
				$\begin{array}{c}
				\mbox{\normalsize $\left\langle \frac{\texttt{let}\ \texttt{mut} \ X \ : \ T;}{.} \right\rangle _{\sf k}$ }
				\left\{ \begin{array}{lll}
				\left\langle  X \Rightarrow L  \right\rangle _{\sf env}& \left\langle  L \Rightarrow \ \perp  \right\rangle _{\sf store} & \left\langle  L \Rightarrow T  \right\rangle _{\sf typeEnv} \\
				\left\langle  L \Rightarrow 1  \right\rangle _{\sf mutType} &  \left\langle  L \Rightarrow \ \perp  \right\rangle _{\sf borrow} & \left\langle  L \Rightarrow \ \perp  \right\rangle _{\sf ref} \\
				\left\langle  L \Rightarrow \ \perp  \right\rangle _{\sf refType} &
				\left\langle  L \Rightarrow \ 0  \right\rangle _{\sf moved}&
				\z L + 1\y_{\sf nextLoc}
				\end{array} \right\} \\
				\hfill \klabel{Declaration-of-Mutable-Variable}
				\end{array}$  \\    \smallskip
				$\begin{array}{cc}
				\mbox{\normalsize $\left\langle \frac{X}{V}  \right\rangle _{{\sf k}}$}
				\left\{ \begin{array}{l}
				\left\langle X \mapsto L  \right\rangle _{{\sf env}} \\  \left\langle  L\mapsto V  \right\rangle _{{\sf store}}
				\end{array} \right\}  \\
				\hfill \klabel{Lookup-of-Variable}
				\end{array}$
				
				$\begin{array}{c}
				\mbox{\normalsize $\left\langle \frac{ X = V;}{.} \right\rangle _{\sf k}$}
				\left\{ \begin{array}{ll}
				\left\langle  X \mapsto L  \right\rangle _{\sf env} & \left\langle  L \mapsto T   \right\rangle _{\sf typeEnv}  \\
				\left\langle  L \mapsto 1 \right\rangle _{\sf mutType} & \left\langle  L \mapsto ( \_ \Rightarrow V  ) \right\rangle _{\sf store}
				\end{array} \right\} \\
				\quad  \qquad \qquad \text{when} \ T =  getType(V) \hfill \klabel{Assignment}
				\end{array}$    \quad
				$\begin{array}{c}
				\mbox{\normalsize $\left\langle \frac{\texttt{let} \ \texttt{mut} \  X:\ [ T;N];}{.}  \right\rangle _{\sf k}$}
				\left\{ \begin{array}{ll}
				\z X \Rightarrow L \y_{\sf env} & \z L \Rightarrow [T;N] \y_{\sf typeEnv}\\
				\z L \Rightarrow 1  \y_{\sf mutType} & \z L+N \y_{\sf nextLoc}  \\
				\multicolumn{2}{l}{\z  L \ldots L+N-1 \Rightarrow \perp \y_{\sf store}}
				\end{array} \right\} \\
				\hfill \klabel{Declaration-of-Mutable-Array}
				\end{array} $
				\\\smallskip
				
				$\begin{array}{r}
				\mbox{\normalsize $\left\langle \frac{X[I]\ =\ V;}{.} \right\rangle_{\sf k}$}
				\left\{\begin{array}{ll}
				\z X \mapsto L \y_{\sf env} & \z L \mapsto [T;N] \y_{\sf typeEnv} \\
				\z L \mapsto 1 \y_{\sf mutType} & \z L+I \mapsto (\_ \Rightarrow V) \y_{\sf store}
				\end{array} \right\} \\
				\mbox{when} \ 0\leq I < N \ \text{and} \ T = getType(V) \ \hfill \klabel{Updating-of-Array-Element}
				\end{array}$
				\hfill
				$\begin{array}{c}
				\mbox{\normalsize $\left\langle\frac{X[I]}{V}  \right\rangle _{\sf k}$}
				\left\{\begin{array}{ll}
				\z X \mapsto L \y_{\sf env} &
				\z L \mapsto [T;N] \y_{\sf typeEnv} \\
				\z L+I \mapsto V  \y_{\sf store}
				\end{array}\right\}\\
				\mbox{when} \ 0\leq I < N  \hfill \klabel{Evaluation-of-Array-Element}
				\end{array}$
				
				\\\hline

				\textbf{Borrow, reference, lifetime and dereference} \\
				$\begin{array}{c}
				\mbox{\normalsize $\left\langle \frac{ X\ =\ \&\text{mut}\ Y;}{.} \right\rangle _{\sf k}$}
				\left\{ \begin{array}{ll}
				\left\langle  X \mapsto L_1 , Y \mapsto L_2 \right\rangle_{\sf env} & \left\langle L_1 \mapsto 1, L_2 \mapsto 1 \right\rangle _{\sf mutType}\\
				\left\langle L_1 \mapsto 0, L_2 \mapsto 0 \right\rangle_{\sf moved} & \left\langle  L_1 \mapsto (\_ \Rightarrow L_2) \right\rangle _{\sf ref}\\
				\left\langle  L_1 \mapsto ( \_ \Rightarrow 1  ) \right\rangle _{\sf refType} & \left\langle  L_2 \mapsto ( \perp \Rightarrow 1  ) \right\rangle _{\sf borrow}
				\end{array} \right\}\\
				\qquad \qquad \qquad \qquad \text{when} \ L_1 > L_2  \hfill \klabel{Mutable-Reference}
				\end{array}$
				\hfill
				
				$\begin{array}{c}
				\mbox{\normalsize $\left\langle \frac{ X\ =\ \& Y;}{.} \right\rangle _{\sf k}$}
				\left\{ \begin{array}{ll}
				\left\langle  X \mapsto L_1 , Y \mapsto L_2 \right\rangle_{\sf env} & \left\langle L_1 \mapsto 1, L_2 \mapsto 1 \right\rangle _{\sf mutType}\\
				\left\langle L_1 \mapsto 0, L_2 \mapsto 0 \right\rangle_{\sf moved} & \left\langle  L_1 \mapsto (\_ \Rightarrow L_2) \right\rangle _{\sf ref}\\
				\left\langle  L_1 \mapsto  0  \right\rangle _{\sf refType} & \left\langle  L_2 \mapsto 0 \right\rangle _{\sf borrow}
				\end{array} \right\}\\
				\qquad \qquad \qquad \qquad \text{when} \ L_1 > L_2  \hfill \klabel{Immutable-Reference} \smallskip
				\end{array}$ \\
				
				$\begin{array}{c}
				\mbox{\normalsize $\left\langle \frac{*X}{V} \right\rangle _{\sf k}$}
				\left\{\begin{array}{ll}
				\left\langle  X \mapsto L_1  \right\rangle _{\sf env} &
				\left\langle  L_1 \mapsto L_2  \right\rangle _{\sf ref} \\
				\left\langle L_1 \mapsto 0 \right\rangle_{\sf moved} &
				\left\langle  L_2 \mapsto V  \right\rangle _{\sf store}
				\end{array}\right\}\\
				\hfill \klabel{Dereference}
				\end{array}$
				
				\\\hline

				%
				
				\textbf{Function definition and function call} \\
				
				$\begin{array}{c}
				\mbox{\normalsize $\left\langle \frac{\texttt{fn}\ F({\tt TIDs})\ \text{-$>$} {\tt T} \{\tt S\}}{.} \right\rangle_{\sf k}$}
				\left\{\begin{array}{ll}
				\left\langle  F \Rightarrow L \right\rangle_{\sf env} \z L+1 \y_{\sf nextLoc} \\
				\left\langle  L \Rightarrow \lambda({\tt TIDs},{\tt S},{\tt T}) \right\rangle_{\sf store}
				\end{array}\right\}\\
				\hfill \klabel{Function-Definition-with-Return-Type}
				\end{array}$
				
				$\begin{array}{c}
				\mbox{\normalsize $\z \frac{{\tt mkDecls}((X\ :\ T , \ {\tt TIDs}), \ (V, {\tt Vs}))}{\texttt{let} \ X\ :\ T = V;  \ {\tt mkDecls}( {\tt TIDs}, \ {\tt Vs})}  \y_{\sf k}$}\\
				\hfill \klabel{mkDecls-Helper-Function}
				\end{array} $
				
				$\begin{array}{c}
				\mbox{\normalsize $\left\langle
					\frac{\lambda({\tt TIDs},{\tt S},{\tt T})({\tt Vs}) \rightsquigarrow K}{{\tt mkDecls}({\tt TIDs},{\tt Vs}) \rightsquigarrow {\tt S} \rightsquigarrow\texttt{return} \ ();} \right\rangle_{\sf k}$ }
				\left\{\begin{array}{ll}
				\langle{(\tt Env},K,{\tt T})\rangle_{\sf fstack} \\
				\langle {\tt Env} \leftarrow \perp \rangle_{\sf env}
				\end{array}\right\}\\
				\hfill \klabel{Function-Call}
				\end{array}$
				\\ \smallskip
				$\begin{array}{r}
				\mbox{\normalsize $\left\langle\frac{\texttt{fn}\ F({\tt TIDs})\ \{\tt S\}}{\texttt{fn}\ F({\tt TIDs})\ \text{-$>$} \ {\tt ()} \ \{\tt S\}} \right\rangle_{\sf k}$}\\
				\hfill \klabel{Function-Definition-without-Return-Type}
				\end{array}$
				\quad \quad \quad
				$\begin{array}{r}
				\mbox{\normalsize $\left\langle\frac{\texttt{fn}\ F({\tt TIDs})\ \text{-$>$} \ {\tt T} \ \{\tt S \ E\}}{\texttt{fn}\ F({\tt TIDs})\ \text{-$>$} \ {\tt T} \ \{\tt \tt S \ return \ E;\}} \right\rangle_{\sf k}$}\\
				\hfill \klabel{Function-Definition-Return-by-Last-Expression}
				\end{array}$
				\quad \quad \quad
				$\begin{array}{c}
				\mbox{\normalsize $\z \frac{\text{return}\ V; \rightsquigarrow - }{V \rightsquigarrow K} \y_{\sf k}$}
				\left\{ \begin{array}{ll}
				\z ({\tt Env},K,{\tt T}) \y_{\sf fstack} \\
				\z \_ \leftarrow {\tt Env} \y_{\sf env}
				\end{array} \right\}\\
				\qquad  \text{when} \ T = getType(V)  \hfill \klabel{Return}
				\end{array} $
				
				\\ \hline
				
				
				\textbf{Struct and ownership} \\
				
				$\begin{array}{c}
				\mbox{\normalsize $\z \frac{{\tt struct} \ Z \ \{\tt TIDs\}}{.} \y_{\sf k}$}
				\left\{ \begin{array}{ll}
				\z Z\Rightarrow L \y_{\sf env} \z L+1 \y_{\sf nextLoc}\\
				\z L \Rightarrow F_1(\tt TIDs)\y_{\sf store} \\
				\end{array} \right\}\\
				\hfill \klabel{Struct-Definition}
				\end{array} $
				
				\hfill
				$\begin{array}{c}
				\mbox{\normalsize $\z \frac{\texttt{let}\ \texttt{mut} \ X\ =\   Z\ \{\tt Vs\};}{F_2(F_1({\tt TIDs}),X,{\tt Vs})} \y_{\sf k}$}
				\left\{ \begin{array}{lll}
				\z X \Rightarrow L_1, \ Z \mapsto L_2 \y_{\sf env} & \z L_1 \Rightarrow 1 \y_{\sf mutType} & \z L_1 \Rightarrow 0 \y_{\sf moved} \\
				\z L_1 \Rightarrow Z \y_{\sf typeEnv} & \z L_2 \mapsto F_1({\tt TIDs}) \y_{\sf store} & \z L+1 \y_{\sf nextLoc}
				\end{array} \right\}\\
				\hfill \klabel{Declaration-of-Struct-Instance}
				\end{array} $ \\

				$\begin{array}{c}
				\mbox{\normalsize $\z \frac{F_2(F_1((Y\ :\ T,\ {\tt TIDs}), \ X, \ (Y\ :\ V), \ {\tt Vs})}{F_2(F_1({\tt TIDs}),  \ X, \ {\tt Vs})} \y_{\sf k}$}
				\left\{ \begin{array}{l}
				\z X.Y \Rightarrow L \y_{\sf env}\\
				\z L \Rightarrow V \y_{\sf store}\\
				\z L + 1\y_{\sf nextLoc}
				\end{array} \right\}\\
				\hfill \klabel{$F_1$-$F_2$-Helper-Function}
				\end{array} $\smallskip
				
				\hfill
				
				$\begin{array}{c}
				\mbox{\normalsize $\z \frac{ X\ = \ Y;} {F_3(X,\ Y,\ {\tt TIDs}) \rightsquigarrow F_4(Y)} \y_{\sf k}$}
				\left\{ \begin{array}{ll}
				\z X \mapsto L_1 , Y \mapsto L_2, Z \mapsto L_3 \y_{\sf env} & \z L_1 \mapsto Z, L _2\mapsto Z \y_{\sf typeEnv}  \\
				\z L_1 \mapsto 1 \y_{\sf mutType}&
				\z L_3 \mapsto F_1({\tt TIDs}) \y_{\sf store}
				\end{array} \right\}\\
				\hfill \klabel{Ownership-Move}
				\end{array} $ \\
				
				$\begin{array}{c}
				\mbox{\normalsize $\z \frac{F_3(X,Y,((P : T), {\tt TIDs}))}{X.P\ =\ Y.P;\ \rightsquigarrow \ F_3(X,Y,{\tt TIDs})} \y_{\sf k}$}
				\\
				\hfill \klabel{$F_3$-Helper-Function}
				\end{array} $\smallskip
				\quad \quad \quad \quad \quad \quad \quad \quad \quad
				
				$\begin{array}{c}
				\mbox{\normalsize $\z \frac{F_4(X)}{.} \y_{\sf k}$}
				\left\{
				\begin{array}{l}
				\z X \mapsto L \y_{\sf env} \\
				\z L \mapsto (0 \Rightarrow 1 ) \y_{\sf moved}
				\end{array}
				\right\}\\
				\hfill \klabel{$F_4$-Helper-Function}
				\end{array}$
				
				\hfill
				$\begin{array}{c}
				\mbox{\normalsize $\z \frac{X.Y}{V}  \y_{\sf k}$}
				\left\{ \begin{array}{ll}
				\multicolumn{2}{l}{\z X.Y \mapsto L_1, \ X \mapsto L_2 \y_{\sf env} }\\
				\z L_1 \mapsto V \y_{\sf store} & \z L_2 \mapsto 0 \y_{\sf moved}
				\end{array} \right\}\\
				\hfill \klabel{Evaluation-of-Struct-Field}
				\end{array} $ \\
				
				\hfill
				$\begin{array}{c}
				\mbox{\normalsize $\z \frac{X.Y\ =\ V;}{.}  \y_{\sf k}$}
				\left\{ \begin{array}{ll}
				\z X.Y \mapsto L_1, X \mapsto L_2 \y_{\sf env}
				&\z L_2 \mapsto 1 \y_{\sf mutType} \\
				\z L_1 \mapsto (\_ \Rightarrow V) \y_{\sf store}
				&\z L_2 \mapsto 0 \y_{\sf moved}
				\end{array} \right\}\\
				\hfill \klabel{Updating-of-Struct-Field}
				\end{array} $
				\\  \hline
		\end{tabular}}
	\end{center}	
	\label{tab:semantics}
\end{table*}

The cell {\sf T} is the top one in $\K$ which contains all the cells.  As aforementioned, the {\sf k} cell contains the computations of a program.
The {\sf env} cell is the local environment, recording the map from variables to their locations.
Inside {\sf control} cell, there is a {\sf fstack} cell which encodes the stack frame. The {\sf fstack} cell is a list, in which each element contains an {\sf env} cell,
some computations and a return type. The {\sf genv} cell represents global environment.
The {\sf typeEnv} cell records the type of a given variable's location. 
The values of all the defined variables are stored in the {\sf store} cell.

Since a variable is either mutable or immutable in Rust, we add {\sf mutType} cell to record whether the variable is mutable or immutable.
When a new variable is declared, a new location that is an integer value is allocated from {\sf nextLoc} cell. After that, integer value in {\sf nextLoc} is increased by one.
The {\sf borrow} cell keeps the record whether a variable is mutably or immutably borrowed if there exists an alive reference to it.
The {\sf ref} cell records reference relations.
The {\sf refType} cell contains the types of references, including immutable and mutable references.
The {\sf moved} cell is a map from locations of variables to $\{0,1\}$, where $1$ denotes that the variable has been moved,
otherwise not.

\subsection{Formalization of the semantics}

In this section, we present the formal semantics $\krust$ in $\K$, emphasizing on key features of Rust such as: mutability, borrows and ownership.
Table \ref{tab:semantics} shows the partial semantic rules of $\krust$, which
specify the semantics of Rust statements by modifying the content of relevant cells.
In the semantic rules, operator $\Rightarrow$ is used to add a new pair to its corresponding cell. For instance,  $\left\langle  X \Rightarrow L  \right\rangle _{\sf env}$ means inserting a new pair $X : L$ into the {\sf env} cell.
The operator $\perp$ denotes \emph{undefined value}.
Dot $\cdot$ is a special character in $\K$, representing the identity element in list and map structures.
Note that curly braces are used only for the compactness of the rules, where we leave out all the dots ``$\cdots$'' in cells.
The complete semantic rules of $\krust$ can be found in its source code.

\subsubsection{Variable declaration and assignment}


The semantic rule \klabel{Declaration-of-Immutable-Variable}
specifies how cells are affected after the execution of the statement $\texttt{let} \ X: T;$ in the {\sf k} cell, where $X$ is a variable and $T$ is a primitive type.
A new pair $L :\ \perp$ is added to {\sf store}, where $\perp$ indicates that $X$ is uninitialized.
The pair  $L:0$ is added to {\sf mutType}, where 0 means that $X$ is \emph{immutable}. The next available location becomes $L+1$ since $X$ has consumed the location $L$.
Some other  initialization operations are made in the relevant cells as shown in the rule. The rule for the declaration of a mutable integer variable is defined likewise with a modification of the {\sf mutType} cell.

The semantic rule \klabel{Assignment} is defined for the ordinary assignment, e.g., assigning a new value to an integer variable.
It says that the assignment could be successfully executed only if $X$ is mutable and the type of $V$ is $T$, where $getType$ is a pre-defined function to get the type of $V$.
The execution updates the value in {\sf store}
using $L \mapsto( \_ \Rightarrow V  )$,  where $\_$ denotes that the current value of $L$ can be any.
The notation $A \mapsto (B \Rightarrow C)$ will be used regularly in this work, which
stands for replacing the pair $A:B$ by $A:C$ in a cell.
A declaration with an initial value can be handled similarly by combining the above semantic rules.

The semantic rule \klabel{Declaration-of-Mutable-Array} is defined for the declaration of a mutable array, where
a new pair $X : L$ is added to {\sf env}, i.e., allocates a location $L$ for $X$.
In {\sf typeEnv}, $[T;N]$ is the array type.
$L \ldots L+N-1$ in {\sf store} denotes the locations $L,\cdots,L+N-1$ which are uninitialized.
The integer in {\sf nextLoc} cell is increased by $N$, as locations $L,\cdots,L+N-1$
are used to store the content of the array. The rules for the evaluation and updating of an array are defined likewise, as depicted in the table.

\subsubsection{Borrowing, reference, lifetime and dereference}
Both \emph{immutable references} and \emph{mutable references} for borrowing are implemented in $\krust$.
\klabel{Mutable-Reference} and \klabel{Immutable-Reference} are two of semantic rules for immutable and mutable references respectively.
The \klabel{Mutable-Reference} rule expresses that if both $X$ and $Y$ are \emph{mutable} and have not been moved,
and $Y$ has not been borrowed yet, then
the value of $L_1$ in {\sf ref} cell is updated to $L_2$, which may be used for dereference,
the reference type of $L_1$ in {\sf refType} is assigned by $1$ denoting that
the reference type of $L_1$ is \emph{mutable},
the pair $L_2 : \perp$ in cell {\sf borrow} is updated to $L_2  : 1$ meaning that $Y$ is now mutably borrowed,
while the condition $L_1 > L_2$ ensures that $X$ has to be declared later than $Y$, i.e., the lifetime of $X$ is included in the lifetime of $Y$.
The \klabel{Immutable-Reference} rule is defined likewise, except that $Y$ is already immutably borrowed and
the reference type of $L_1$ in {\sf refType} is assigned by $0$ denoting
\emph{immutable}.
The intuition behind the \klabel{Dereference} rule is straightforward, namely,
the value of $*X$ is evaluated via the reference relation $ L_1 \mapsto L_2 $ in the {\sf ref} cell.

\subsubsection{Function definition and function call}

The standard function definition that has an explicit return type is handled by the \klabel{Function-Definition-with-Return-Type} rule,
which allocates a location $L$ for the name $F$ in the {\sf env} cell and binds the location $L$ with an auxiliary function $\lambda()$ in {\sf store}.
The auxiliary function $\lambda()$ records the type and body of the function, which will be used for function call, i.e., the \klabel{Function-Call} rule.

The function call is processed into two steps: first replacing the function name by its $\lambda()$ function in {\sf store} using the \klabel{Lookup-of-Variable} rule,
then applying the \klabel{Function-Call} rule. The \klabel{Function-Call} rule first stores the current local environment ${\tt Env}$ together with
the return type ${\tt T}$ and the remaining computations $K$ into the ${\sf fstack}$ cell, reallocates a new empty local environment denoted by $\z {\tt Env} \leftarrow \ \perp \y_{\sf env}$
in the rule, and sets ${\tt mkDecls}({\tt TIDs},{\tt Vs}) \rightsquigarrow {\tt S} \rightsquigarrow\texttt{return}\ ();$ as the computations.
The latter firsts executes the helper function ${\tt mkDecls}()$ which is used to
declare formal parameters initialized with actual parameters ${\tt Vs}$ recursively (c.f. the \klabel{mkDecls-Helper-Function} rule), then executes
the function body ${\tt S}$ followed by an additional return statement $\texttt{return}\ ();$.
We remark that  \texttt{return ();} is added to handle the case
that the function definition does not have a return statement, and it will be executed only if there is no \texttt{return} statement at the end of the function body ${\tt S}$.

The semantic rule \klabel{Function-Definition-without-Return-Type} handles the corner case that the function definition does not have a return type,
for which, the unit type $()$ is added as the return type.
The semantic rule \klabel{Function-Definition-Return-by-Last-Expression} handles  another corner case that the function definition returns the value of the last expression,
for which, the last expression ${\tt E}$ is rewritten as a return statement $\texttt{return\ E};$.

The  rule \klabel{Return} first checks the type of the return value $V$, then returns the value $V$ and finally restores the local environment {\tt Env} and the computations $K$ from the top of the {\sf fstack} cell at the same time.

\subsubsection{Struct and ownership}
The rule  \klabel{Struct-Definition}  is used for struct definition, which adds a new pair $Z:L$
into the ${\sf env}$ cell. The value of $L$ in ${\sf store}$ is a helper function $F_1$ that is used to record the fields of the struct.

The rule \klabel{Declaration-of-Struct-Instance} is defined for the declaration of a mutable struct instance.
It allocates a location $L_1$ for $X$ (i.e., adds $X:L_1$ into {\sf env}) and sets the type of $L_1$ as $Z$, which is a struct name (i.e., add $L_1:Z$ into ${\sf typeEnv}$).
Other related cells are initialized accordingly.
The declaration and initialization of fields are handled by the computation
$F_2(F_1({\tt TIDs}),X,{\tt Vs})$, which is defined by the \klabel{$F_1$-$F_2$-Helper-Function} rule.
This rule recursively allocates a location for each field, which is initialized with the corresponding value from {\tt Vs}.

The rule \klabel{Ownership-Move} specifies Rust's move semantics, in which
the assignment statement is encoded as the computations of two helper functions $F_3$ and $F_4$,  if $X$ and $Y$ have same type and $X$ is immutable.
Notice that the pair $L_3 : F_1({\tt TIDs})$ in {\sf store} ensures that $Y$ is a struct instance.
The semantics of the helper functions $F_3$ and $F_4$ are expressed by the rules
\klabel{$F_3$-Helper-Function} and \klabel{$F_4$-Helper-Function} respectively.
Intuitively, \klabel{$F_3$-Helper-Function} helps to copy the fields from $Y$ to $X$. \klabel{$F_4$-Helper-Function} is used to update the {\sf move} cell.
We remark that $\z L_1 \mapsto 0, L_2 \mapsto 0 \y_{\sf moved}$ is not added in the \klabel{Ownership-Move} rule, as ``$X.P = Y.P;$'' in \klabel{$F_3$-Helper-Function} ensures that both variables are not moved yet.
The semantic rules \klabel{Evaluation-of-Struct-Field} and \klabel{Updating-of-Struct-Field} are defined for
evaluation and updating of a struct filed, in which it is required that the struct variable is not moved, i.e., the pair $L_2: 0$
should occur in {\sf moved}.
\section{Testing and Applications}
\label{sec:test}
In this section, we validate our semantics $\krust$ and show
some potential applications of $\krust$.

\subsection{Conformance Testing}
Following previous work \cite{ER12,FM14,BR15,PSR15} which
used test suite for validating executable language semantics, we tried to do the same.
We validated our semantics $\krust$ by
testing the $\krust$ interpreter (that was automatically generated from the $\krust$ semantics using $\K$ framework)
against both the official test suite of Rust \footnote{\url{https://github.com/rust-lang/rust/tree/master/src/test.}} and hand-crafted tests.

The official test suite of Rust is used to test the Rust compiler.
It is already split into folders containing different categories of tests.
We chose tests from the ``run-pass'' folder, as
others were designed for different purposes such as error message.
There are 3119 tests in the ``run-pass'' folder: some of them can be compiled by the nightly version or stable version of the compiler,
and some may be ignored during compiler testing, but it is unclear how these tests were used from the official documents.
Therefore, we parsed all 3119 and 195 of them are supported by our syntax.
In 195 tests, there are 38 tests that either do not have a \emph{main} function or
call some standard library functions. Therefore, we chose other 157 tests.

Because 157 tests might not cover all the supported constructs,
we hand-crafted 25 tests according to the syntax defined in Table \ref{tab:syntax}.
25 tests together cover all the primitive constructs.

We have tested the $\krust$ interpreter against all these 182 tests.
$\krust$ successfully parsed all of them and the results produced by the interpreter are same as
the one produced by the compiled programs using the Rust compiler.

Remark that the semantic coverage of the test suite has
not been well-studied, we leave this to future work.

\subsection{Applications}
One of the main goals of our semantics is to provide a formal semantics for Rust.
Beyond just giving a formal reference for the defined language,
there are many applications of our formal semantics using the language-independent tools provided by $\K$.
In this work, we demonstrate this by showing two applications:
\emph{debugging} and \emph{verification}, which are automatically derived from the semantics $\krust$.

\noindent
\noindent{\bf Debugging.}
We can turn the $\K$ debugger into a debugger for Rust which allows users to inspect program states.
We demonstrate this by the following example.
\begin{lstlisting}
fn main() {
    let mut x: i32 = 10;
    while x > 0 {
        x = x - 1;
    } 
}	
\end{lstlisting}

We can debug this program using the command.
\begin{lstlisting}
krun test.rs --debugger 	
\end{lstlisting}

Users can step through one or more semantic rules individually from the current point and print the current state.
For instance, after executing Line 4 once, part of the state will look like below:
\begin{lstlisting}
<env> x |-> 1 </env>
<typeEnv> 1 |-> i32 </typeEnv>
<mutType> 1 |-> 1 </mutType>
<store> 1 |-> 9 </store>
\end{lstlisting}
\medskip
\noindent{\bf Verification.}
A \emph{sound-by-construction} program verifier for Rust can be automatically derived from
$\krust$ without additional effort. The verifier allows us to automatically check reachability properties
including all Hoare-style functional correctness claims and time complexity of a computation.
As an example, we will verify the time complexity of the Euclidean algorithm by subtraction which computes greatest common divisor.

\begin{lstlisting}
fn gcd(a: i32, b : i32) -> i32 {
    if a!=b {
        if a>b { return gcd(a-b, b); }
        else   { return gcd(a, b-a); }
    }else { return a; }
}
\end{lstlisting}

The Euclidean algorithm is implemented in Rust as shown above.
We will prove its time complexity is indeed ${\bf O}(\max(a,b))$.
To verify time complexity, an extra cell {\sf time} is added to the configuration which increases a counter $T$ each
time when \texttt{gcd} is called. The core part of the specification for proving is:
\begin{lstlisting}
 gcd(X:Int, Y:Int)
 <time> T1 => T2 </time>
 requires X > 0 , Y > 0
 ensures T2 - T1 <= maxInt(X,Y) 
\end{lstlisting}
where \texttt{T1} and \texttt{T2} respectively denote the value of the counter $T$ in pre-state and post-state of the function,
{\tt requires} and {\tt ensures} respectively denote pre-condition and post-condition,
 {\tt T2 - T1} denotes the number of calls to \texttt{gcd},
and \texttt{maxInt} is a built-in function which returns the larger one of two integers.
The verifier outputs {\tt true} which proves that {\tt T2 - T1 <= maxInt(X,Y)} holds, i.e., the time complexity is ${\bf O}(\max(a,b))$.


\section{Related work}\label{sec:rel}
A multitude of formal semantics for real programming languages have been proposed in the literature.
Due to space restriction, we only discuss large semantics in $\K$ and other works related to Rust.

\subsection{Other formal semantics in $\K$}
Ellison and Rosu defined an executable formal semantics for C11, which has been extensively tested against the GCC torture suite \cite{ER12}
and evaluated by debugging, monitoring, and (LTL) model checking of C programs using the built-in capabilities of $\K$.
Hathhorn et al. defined undefined behavior in C11 \cite{HER15}, complementing the semantics of \cite{ER12}.

Filaretti and Maffeis defined a formal semantics for PHP \cite{FM14}. As there is no official language standard for PHP,
they had to heavily rely on testing against the some test suite.
Their semantics has been evaluated by model checking certain properties of some programs.

Bogdanas and Rosu \cite{BR15} gave a formal semantics for Java 1.4, which is split into two phases: a static semantics and a dynamic semantic.
The static semantics enriches the original Java program by annotating statically inferred information, while the dynamic one gives the executable semantics.
Their semantics has been evaluated by model checking multi-threaded programs.

Park et al. \cite{PSR15} presented a formal semantics for JavaScript which has been tested against the ECMAScript 5.1 conformance test and passes all core language tests.

Besides the above languages, the formal semantics of Python 3.3, Verilog, Scheme, LLVM IR and Esolangk were also defined in $\K$ \cite{gut13,MKMR10,MHR07,llvmk,Esolangk}.

Compared to other language semantics in $\K$, the most distinguished aspect of our semantics is the formalization of
the distinct features of Rust, namely \emph{ownership}, \emph{borrow}, \emph{lifetime}, etc.
To address these features,  we had to redesign the \emph{program state} instead of just copying from the existing works.
Our semantics has been validated using the Rust standard test suite as well as hand-crafted tests.

\subsection{Works related to Rust}
Rust has attracted some attention of researchers, and there are some works
on the Rust type system and Rust program verification.

Reed presented a formal semantics for Rust that captures the key features relevant to memory safety, unique pointers and borrowed references, described the operation of the Borrow Checker,
and formalized the Rust type system \cite{Ree15}. The main goal of this work is to provide a framework for borrow checking.
However, Rust has been evolved a lot since their work and their semantics has not been implemented yet, hence not executable.
Very recently, Jung et al. defined a formal semantics and type system of a language $\lambda_{\text{Rust}}$, which incorporates Rust's notions \cite{JKD18}.
Their work has been implemented in Coq.
The main goal of this work is to study Rust's ownership discipline in the presence of unsafe code.
However, the language $\lambda_{\text{Rust}}$ is close to Rust's Mid-level Intermediate Representation (MIR) rather than the actual Rust language,
and their semantics defines more behaviors than Rust does. Our semantics addresses to exact behavior of the actual Rust language and is executable.

Dewey et al. proposed a technique to fuzz type checker implementations and applied to test Rust's type checker \cite{DRH15}.
It has identified 18 bugs. It is evident that formalization of the Rust type system is important, while formalization of the Rust semantics
is the first step toward this.

Two model checkers for verifying unsafe Rust programs have been proposed \cite{TPT15,Hah16}, which verify Rust programs by translating them
into input languages of existing model checkers. Our semantics can be automatically turned into formal analysis tools such as state-space explorer for reachability, model-checker and symbolic execution engine
using the language-independent tools provided by $\K$ \cite{SPYKR16}.

\section{Conclusion, Limitations and Future Work}\label{sec:discu}
In this work, we proposed a formal executable semantics $\krust$ for Rust using the $\K$ framework.
$\krust$ captures (1) all the primitive types and their operations, (2) compound data types: {\tt struct}, {\tt array} and {\tt vector},
(3) all the basic control flow constructs: {\tt for}, {\tt while}, {\tt loop}, {\tt if}, {\tt if/else}, {\tt function definition/call}
and (4) three most distinct and important features: {\tt ownership}, {\tt borrow} and {\tt lifetime}.
We tested $\rust$ on many hand-crafted tests and Rust's official tests suite.
Tests in supported syntax are all passed. We also demonstrated potential applications of $\krust$ for debugging and verification of Rust
programs.

However, $\krust$ does not cover the full features of Rust, as Rust is being actively developed and some of these features are not stable so far.
As a witness, although the Rust's community provides some syntax in EBNF \cite{Grammar},
it is still far away from complete. This makes the formalization of Rust much difficult, as mentioned in \cite{JKD18}.
The following is a list of features that we haven't implemented yet but plan to implement:
(1) structs with reference fields,
(2) pattern matching which can be seen as a generalization of switch-case,
(3) trait objects which like interfaces in Java,
(4) lifetime annotation which is used to mark explicit lifetime in functions or structs,
(5) complex closures which use outside variables,
(6) concurrency for writing multi-threading programs,
(7) crates and modules which are used to call external library codes
(8) unsafe which is used to write code that the Rust compiler is unable to prove its safety,
etc.
A long-term program is to develop an almost complete formal executable semantics for Rust and formally verify Rust programs using formal analysis tools turned from the semantics,
towards which the work reported in this paper is the first cornerstone.

\balance
\bibliographystyle{IEEEtran}
\bibliography{main}

\begin{thebibliography}{10}
\providecommand{\url}[1]{#1}
\csname url@samestyle\endcsname
\providecommand{\newblock}{\relax}
\providecommand{\bibinfo}[2]{#2}
\providecommand{\BIBentrySTDinterwordspacing}{\spaceskip=0pt\relax}
\providecommand{\BIBentryALTinterwordstretchfactor}{4}
\providecommand{\BIBentryALTinterwordspacing}{\spaceskip=\fontdimen2\font plus
\BIBentryALTinterwordstretchfactor\fontdimen3\font minus
  \fontdimen4\font\relax}
\providecommand{\BIBforeignlanguage}[2]{{%
\expandafter\ifx\csname l@#1\endcsname\relax
\typeout{** WARNING: IEEEtran.bst: No hyphenation pattern has been}%
\typeout{** loaded for the language `#1'. Using the pattern for}%
\typeout{** the default language instead.}%
\else
\language=\csname l@#1\endcsname
\fi
#2}}
\providecommand{\BIBdecl}{\relax}
\BIBdecl

\bibitem{matsakis2014rust}
N.~D. Matsakis and F.~S. Klock~II, ``The {Rust} language,'' in \emph{ACM SIGAda
  Ada Letters}, vol.~34, no.~3, 2014, pp. 103--104.

\bibitem{Redox}
Redox, ``Redox: a unix-like operating system written in rust,''
  https://www.redox-os.org, 2018.

\bibitem{ABGMMMS16}
B.~Anderson, L.~Bergstrom, M.~Goregaokar, J.~Matthews, K.~McAllister,
  J.~Moffitt, and S.~Sapin, ``Engineering the servo web browser engine using
  rust,'' in \emph{{ICSE'16}}, 2016, pp. 81--89.

\bibitem{DDLCZCWW17}
Y.~Ding, R.~Duan, L.~Li, Y.~Cheng, Y.~Zhang, T.~Chen, T.~Wei, and H.~Wang,
  ``{POSTER:} rust {SGX} {SDK:} towards memory safety in intel {SGX} enclave,''
  in \emph{{CCS'17}}, 2017, pp. 2491--2493.

\bibitem{Stackoverflow}
Stackoverflow, ``https://stackoverflow.com/search?q=rust+,'' 2018.

\bibitem{ro2010}
G.~Ro{\c{s}}u and T.~F. {\c{S}}erb{\v{a}}nu{\v{t}}{\v{a}}, ``An overview of the
  k semantic framework,'' \emph{The Journal of Logic and Algebraic
  Programming}, vol.~79, no.~6, pp. 397--434, 2010.

\bibitem{ER12}
C.~Ellison and G.~Rosu, ``An executable formal semantics of {C} with
  applications,'' in \emph{{POPL'12}}, 2012, pp. 533--544.

\bibitem{BR15}
D.~Bogdanas and G.~Ro{\c{s}}u, ``{K-Java}: a complete semantics of {Java},'' in
  \emph{{POPL'15}}, 2015, pp. 445--456.

\bibitem{SPYKR16}
A.~\c{S}tef\u{a}nescu, D.~Park, S.~Yuwen, Y.~Li, and G.~Ro\c{s}u,
  ``Semantics-based program verifiers for all languages,'' in
  \emph{OOPSLA'16}.\hskip 1em plus 0.5em minus 0.4em\relax ACM, 2016, pp.
  74--91.

\bibitem{gut13}
D.~Guth, ``A formal semantics of {Python} 3.3,'' Master's thesis, University of
  Illinois at Urbana-Champaign, 2013.

\bibitem{FM14}
D.~Filaretti and S.~Maffeis, ``An executable formal semantics of {PHP},'' in
  \emph{{ECOOP'14}}, 2014, pp. 567--592.

\bibitem{Grammar}
\url{https://doc.rust-lang.org/grammar.html}.

\bibitem{PSR15}
D.~Park, A.~Stefanescu, and G.~Rosu, ``{KJS:} a complete formal semantics of
  javascript,'' in \emph{{PLDI'15}}, 2015, pp. 346--356.

\bibitem{HER15}
C.~Hathhorn, C.~Ellison, and G.~Rosu, ``Defining the undefinedness of {C},'' in
  \emph{{PLDI'15}}, 2015, pp. 336--345.

\bibitem{MKMR10}
P.~O. Meredith, M.~Katelman, J.~Meseguer, and G.~Rosu, ``A formal executable
  semantics of {Verilog},'' in \emph{{MEMOCODE'10}}, 2010, pp. 179--188.

\bibitem{MHR07}
P.~Meredith, M.~Hills, and G.~Rosu, ``A {K} definition of {Scheme},''
  University of Illinois at Urbana-Champaign, Tech. Rep., 2007.

\bibitem{llvmk}
{LLVM IR in K}, ``http://github.com/davidlazar/llvm-semantics.''

\bibitem{Esolangk}
E.~in~{K}, ``http://esolang-semantics.googlecode.com.''

\bibitem{Ree15}
E.~Reed, ``Patina: A formalization of the rust programming language,''
  University of Washington, Tech. Rep., 2015.

\bibitem{JKD18}
R.~Jung, J.~Jourdan, R.~Krebbers, and D.~Dreyer, ``{RustBelt}: securing the
  foundations of the {Rust} programming language,'' \emph{{PACMPL}}, vol.~2,
  no. {POPL}, pp. 66:1--66:34, 2018.

\bibitem{DRH15}
K.~Dewey, J.~Roesch, and B.~Hardekopf, ``Fuzzing the {Rust} typechecker using
  {CLP} {(T)},'' in \emph{{ASE'15}}, 2015, pp. 482--493.

\bibitem{TPT15}
J.~Toman, S.~Pernsteiner, and E.~Torlak, ``Crust: {A} bounded verifier for
  {Rust},'' in \emph{{ASE'15}}, 2015, pp. 75--80.

\bibitem{Hah16}
F.~Hahn, ``Rust2viper: Building a static verifier for rust,'' Master's thesis,
  ETH Z\"{u}rich, 2016.

\end{thebibliography}

\end{document}